\documentclass[aps,prb,twocolumn,superscriptaddress]{revtex4-2}

\usepackage{dsfont}
\usepackage{graphicx}			
\usepackage{dcolumn}			
\usepackage{bm}					
\usepackage{hyperref}			

\usepackage{amsmath}
\usepackage{amsfonts}
\usepackage{amssymb}
\usepackage{pstcol,times,xcolor,graphicx,graphics,pstricks,pst-node,pst-coil,pst-grad}
\usepackage[normalem]{ulem}
\usepackage{bm}

\begin{document}


\title{Tuning resonance energy transfer with magneto-optical properties of graphene} 


\author{P. P. Abrantes}
\email{patricia@pos.if.ufrj.br}
\affiliation{Instituto de F\'{\i}sica, Centro de Tecnologia, Universidade Federal do Rio de Janeiro, 21941-972, RJ, Brazil}

\author{G. Bastos}
\email{gbstravassos@gmail.com}
\affiliation{Modelling Methods in Engineering and Geophysics Laboratory (LAMEMO), Departamento de Engenharia Civil, COPPE,  Universidade Federal do Rio de Janeiro, 21941-596, RJ, Brazil}

\author{D. Szilard}
\email{daniela@if.ufrj.br}
\affiliation{Instituto de F\'{\i}sica, Centro de Tecnologia, Universidade Federal do Rio de Janeiro, 21941-972, RJ, Brazil}

\author{C. Farina}
\email{farina@if.ufrj.br}
\affiliation{Instituto de F\'{\i}sica, Centro de Tecnologia, Universidade Federal do Rio de Janeiro, 21941-972, RJ, Brazil}

\author{F. S. S. Rosa}
\email{frosa@if.ufrj.br}
\affiliation{Instituto de F\'{\i}sica, Centro de Tecnologia, Universidade Federal do Rio de Janeiro, 21941-972, RJ, Brazil}



\begin{abstract}

We investigate the resonance energy transfer (RET) rate between two quantum emitters near a suspended graphene sheet in vacuum under the influence of an external magnetic field. We perform the analysis for low and room temperatures and show that, due to the extraordinary magneto-optical response of graphene, it allows for an active control and tunability of the RET even in the case of room temperature. We also demonstrate that the RET rate is extremely sensitive to small variations of the applied magnetic field, and can be tuned up to a striking six orders of magnitude for quite realistic values of magnetic field. Moreover, we evidence the fundamental role played by the magnetoplasmon polaritons supported by the graphene monolayer as the dominant channel for the RET within a certain distance range. Our results suggest that magneto-optical media may take the manipulation of energy transfer between quantum emitters to a whole new level, and broaden even more its great spectrum of applications.

\end{abstract}

\maketitle

\pagebreak


\section{Introduction}


Resonance energy transfer (RET) \cite{Forster1946,GovorovBook,Jones2019} constitutes an important mechanism through which an excited quantum emitter (donor) may transfer its energy to a neighboring one in the ground state (acceptor). Amid the several situations where RET plays a relevant role, a remarkable example is the light harvesting process in plants, in which chlorophyll molecules are excited by the absorption of light and can efficiently transfer this excitation energy to their neighboring molecules \cite{Scholes2011,Bredas2016}.

Different energy transfer mechanisms have been extensively discussed not only in physics, but also in several areas like chemistry, biology and engineering. An efficient energy transfer allows for a variety of applications, such as photovoltaics \cite{Chanyawadee2009}, luminescence \cite{Baldo2000,Song2020}, sensing \cite{Diaz2018}, quantum information \cite{Unold2005,Argyropoulos2019}, and many others. Due to these numerous applications and to advances in different areas combined with the great development of new technologies, controlled modification of the RET rate has also become a topic of huge interest. In this context, substantial theoretical and experimental efforts have been dedicated to investigate the influence of different geometries and materials, such as planar geometries \cite{Marocico2011,Poddubny2015,Bouchet2016}, cavities \cite{Andrew2000,Ghenuche2014}, nanoparticles \cite{Xie2009,Vincent2011,Aissoui2017,Vetrone2018,Schatz2018,Bohlen2019}, cylinders \cite{Marocico2009,Karanikolas2014} and waveguides \cite{Argyropoulos2019,Marocico2011,Marticano2010,deRoque2015,Fiscelli2018}.

Among the progress in so many areas, the field of plasmonics stands out with intense growth in recent decades. Plasmonics consists in the study of the science and applications of the surface plasmon polaritons, which are electromagnetic surface waves coupled to the conduction electrons to form collective charge excitations that propagate at the interface between a dielectric and a conductor \cite{MaierBook,NunoBook}. In particular, surface plasmons supported by graphene are confined much more strongly and present longer propagation lengths when compared to those in conventional noble metals \cite{NunoBook,Iranzo2018,Ni2018}. Another important advantage is their chemical potential tunability that can be achieved by gating and doping \cite{NunoBook,Grigorenko2012,deAbajo2014}. In this sense, graphene provides a suitable platform for manipulation of light-matter interaction and the influence on the RET rate between two emitters has already been analysed both for the case of a monolayer \cite{Velizhanin2012,Biehs2013,Karanikolas2015} and for a nanodisk \cite{Karanikolas2016}. In all of them, the authors explore precisely the change in the RET rate caused by the possibility of tuning the chemical potential.

However, when submitted to an external magnetic field, plasmons and cyclotron excitations hybridize, originating new modes in graphene, named magnetoplasmon polaritons (MPPs) \cite{NunoBook,Ferreira2012}. The MPPs may enhance even more the light-matter interactions, creating a new opportunity to actively control the RET. In this paper we take advantage of graphene's magneto-optical response and propose a setup that takes the degree of RET manipulation to unprecedented levels: two emitters placed in the vicinity of a suspended graphene monolayer in vacuum, submitted to an external magnetic field applied perpendicularly to the monolayer. We demonstrate that the RET rate may change dramatically with respect to the result in free space even for small modulations of the magnetic field. Furthermore, this giant effect may be obtained even for somewhat modest values of the field. Interestingly, our results suggest that magnetoactive materials could act as a logic gate in some practical circumstances, meaning that they could be turned on and off without the need of physical contact, specially at room temperature. Our findings show that a magnetic field applied to the graphene monolayer can be used as an external agent for tuning continuously RET rates. 

This paper is organized as follows. In Sec. \ref{SecGreenFunction} we introduce the system under investigation, the Green's tensor formalism used in the calculation of the RET rate between two emitters in the presence of an arbitrary environment and some important features related to the graphene's response to the external applied magnetic field. In particular, we provide an analysis of how graphene's conductivities vary as a function of the magnetic field, exploring their behavior for distinct values of chemical potential and temperature. Section \ref{SecResults} comprises our main results on the resonance energy transfer between the emitters. For example, we highlight the understanding of the MPPs as the fundamental agents to achieve the intense variations of the RET rate. Section \ref{SecConclusions} is left for final comments and conclusions.


\section{\label{SecGreenFunction}Resonance energy transfer close to a graphene sheet in a magnetic field}

\begin{figure}
\begin{center}
\includegraphics[width=7.8cm]{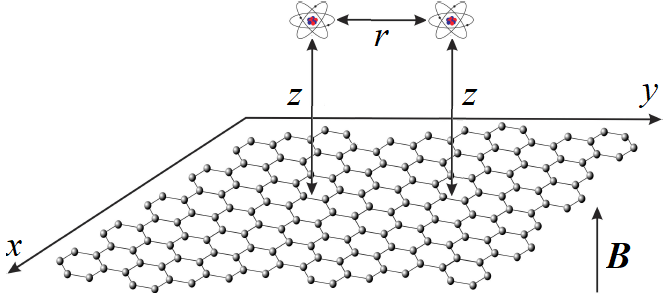}
\end{center}
\vskip -0.6cm
\caption{A pair of two-level emitters separated by a distance $r$, both at a distance $z$ from a suspended graphene sheet. An external magnetic field $\bm{B} = B \bm{\hat{z}}$ is applied perpendicularly to the sheet.}
\label{System}
\end{figure}

In this work we shall be concerned with the RET rate between a pair of two-level quantum emitters $A$ (in the excited state) and $B$ (in the ground state), separated by a distance $r$, both at the same distance $z$ from a suspended graphene sheet in vacuum in thermal equilibrium at temperature $T$. Moreover, the graphene sheet is subjected to a uniform and static external magnetic field $\bm{B} = B \bm{\hat{z}}$ applied perpendicularly to it, as sketched in Fig. \ref{System}.

In the following subsections, we briefly introduce the Green function approach commonly used to calculate the modified RET rate between two quantum emitters when placed in the vicinity of any medium. Then, we move on to the description of the graphene's response to the applied magnetic field, presenting the main equations needed to determine the new RET rate in this particular case.


\subsection{\label{SubsecGreenForm}Methodology}


In the presence of an arbitrary environment, the RET rate $\Gamma$ between two quantum emitters in vacuum located at $\bm{r}_A$ and $\bm{r}_B$, such that $r = |\bm{r}_B - \bm{r}_A|$, normalized by the RET rate in free space $\Gamma^{(0)}$ can be written as \cite{Marocico2009}
\begin{equation}
	\frac{\Gamma}{\Gamma^{(0)}} = \frac{\big\vert \bm{d}_B \cdot \mathds{G} (\bm{r}_B, \bm{r}_A, \omega_0)\cdot \bm{d}_A \big\vert^2}{\big\vert \bm{d}_B \cdot \mathds{G}^{(0)} (\bm{r}_B, \bm{r}_A, \omega_0)\cdot \bm{d}_A \big\vert^2} \,,
\label{RETRate}
\end{equation}

\noindent where $\omega_0$ is the transition frequency of the emitters, $\bm{d}_A$ and $\bm{d}_B$ are their transition electric dipole moments and $\mathds{G}$ and $\mathds{G}^{(0)}$ are the electromagnetic Green dyadics of the full setup and in free space, respectively. The electromagnetic Green dyadic satisfies
\begin{equation}
	\left[ \nabla \! \times \! \nabla \! \times \, - 
\epsilon (\omega, \bm{r}) \frac{\omega^2}{c^2} \right] \mathds{G} (\bm{r}, \bm{r}^\prime, \omega) = - \delta (\bm{r} - \bm{r}^\prime)\, \mathds{I}
\label{EMGDHelmoltz}
\end{equation}

\noindent with the appropriate boundary conditions \cite{NovotnyNanoOptics}, where $c$ is the light velocity in vacuum and $\epsilon (\omega, \bm{r})$ stands for the electric permittivity of the medium. In our case, we take $\epsilon (\omega, \bm{r}) = \epsilon_0$, where $\epsilon_0$ is the electric permittivity of vacuum. It will be convenient to separate the Green dyadic as a sum of two contributions, namely
\begin{equation}
\mathds{G} (\bm{r}_B, \bm{r}_A, \omega_0) = \mathds{G}^{(0)} (\bm{r}_B, \bm{r}_A, \omega_0) + \mathds{G}^{(\textrm{S})} (\bm{r}_B, \bm{r}_A, \omega_0) \,.
\end{equation}

\noindent In this expression $\mathds{G}^{(0)} (\bm{r}_B, \bm{r}_A, \omega_0)$ is the solution to Eq. (\ref{EMGDHelmoltz}) in the absence of any object and $\mathds{G}^{(\textrm{S})} (\bm{r}_B, \bm{r}_A, \omega_0)$ represents the scattered part of the Green function and must obey the electromagnetic field boundary conditions \cite{NovotnyNanoOptics} at the graphene sheet. The procedure to evaluate the scattered part of the total Green function follows from the equation \cite{NovotnyNanoOptics}
\begin{equation}
	\mathds{G}^{(\textrm{S})} = \frac{i}{2} \int \frac{d^2 \bm{k}_{\|}}{\left( 2\pi \right)^2}\, \mathds{R} \, \frac{e^{i\left[ \bm{k}_{\|} \cdot \left( \bm{r}_B - \bm{r}_A \right) + k_{0z} \left( z_B + z_A \right) \right]}}{k_{0z}} \,,
\end{equation}

\noindent where
\begin{equation}
	\mathds{R} = \!\! \sum_{p, q = \{\textrm{TE,TM}\}} \!\! r^{p, q} \, \bm{\epsilon}_{p}^{+} \otimes \bm{\epsilon}_{q}^{-}
\end{equation}

\noindent denotes the reflection matrix with $r^{p, q}$ corresponding to the reflection coefficient for an incoming $q$-polarized wave that is reflected as a $p$-polarized one \cite{NovotnyNanoOptics}. In addition, the TE- and TM-polarization unitary vectors are defined as
\begin{align}
	\bm{\epsilon}_{\textrm{TE}}^{+} &= \bm{\epsilon}_{\textrm{TE}}^{-} = \frac{- k_y \bm{\hat{x}} + k_x \bm{\hat{y}}}{k_{\|}} \,, \\
	\bm{\epsilon}_{\textrm{TM}}^{\pm} &= \frac{\pm k_{0z} (k_x \bm{\hat{x}} + k_y \bm{\hat{y}}) -  k_{\|}^2 \bm{\hat{z}}}{k_{\|} (\omega_0/c)} \,,
\end{align}

\noindent with $\bm{k}_{\|} = k_x \bm{\hat x} + k_y \bm{\hat y}$ and $k_{0z} = \sqrt{(\omega_0/c)^2 - k_{\|}^2}$.

For the sake of simplicity, we analyze emitters with both transition dipole moments being oriented along the $z$-axis (and perpendicular to the graphene sheet), such that Eq. (\ref{RETRate}) reduces to
\begin{equation}
	\frac{\Gamma}{\Gamma^{(0)}} = \frac{\big\vert \mathds{G}_{zz} (\bm{r}_B, \bm{r}_A, \omega_0) \big\vert^2}{\big\vert \mathds{G}^{(0)}_{zz} (\bm{r}_B, \bm{r}_A, \omega_0) \big\vert^2} \,.
\label{RETRateSimpl}
\end{equation}

\noindent More explicitly, we can write \cite{NovotnyNanoOptics}
\begin{equation}
	\mathds{G}^{(0)}_{zz} = \frac{e^{i \omega_0 r/c}}{4 \pi r} \left[ 1- \left( \frac{c}{\omega_0 r} \right)^2 + \frac{i c}{\omega_0 r} \right]
\label{G0zz}
\end{equation}

\noindent and $\mathds{G}^{(\textrm{S})}_{zz} = \bm{\hat z} \cdot \mathds{G}^{(\textrm{S})} \cdot \bm{\hat z}$ is the only contribution of the scattered Green function that needs to be considered, given by
\begin{equation}
	\mathds{G}^{(\textrm{S})}_{zz} = \frac{i c^2}{8 \pi^2 \omega_0^2} \int d\bm{k}_{\|} \frac{k_{\|}^2 \, r^{\textrm{TM,TM}} \, e^{i\left[ \bm{k}_{\|} \cdot \left( \bm{r}_B - \bm{r}_A \right) + k_{0z} \left( z_B + z_A \right) \right]} }{k_{0z}} \,.
\end{equation}

\noindent Writing this equation in polar coordinates, performing the angular integration and identifying $z_A = z_B = z$, we get 
\begin{equation}
	\mathds{G}^{(\textrm{S})}_{zz} =  \frac{i c^2}{4\pi \omega_0^2} \int_0^\infty \!\!\! dk_{\|} \frac{k_{\|}^3 \, J_0 (k_{\|} r) \,r^{\textrm{TM,TM}} \, e^{2i k_{0z} z}}{k_{0z}} \,,
\label{GSzz}
\end{equation}

\noindent where $J_0$ is the cylindrical Bessel function of zeroth order. It is worth mentioning that all information about the influence of the environment is only encoded in $r^{\textrm{TM,TM}}$, which denotes the reflection coefficient of an incoming TM-polarized wave that is reflected with the same TM-polarization \cite{NovotnyNanoOptics}. This arises as a direct consequence of our choice for the direction of the transition dipole moments as being perpendicular to the medium, so they do not couple to TE waves.


\subsection{\label{SubsecGraphProp}Reflection coefficient and conductivities of graphene in a magnetic field}


According to Eq. (\ref{GSzz}), in order to evaluate the scattered Green function, it is required the reflection coefficient $r^{\textrm{TM,TM}}$. It is well known that graphene is a magneto-optical material, in the sense that, under the influence of a perpendicular external magnetic field, its conductivity becomes a tensor with nonzero diagonal and nondiagonal elements and we need to take into account a transverse conductivity ($\sigma_{xy}$), in addition to the standard longitudinal one ($\sigma_{xx}$). The existence of the former contribution makes the TM reflection coefficient slightly more complicated than usual, to wit \cite{Tse2012, KortKamp2014}
\begin{equation}
	r^{\textrm{TM,TM}} = \frac{2 Z^{\textrm{E}} \sigma_{xx} + \eta_0^2 (\sigma_{xx}^2 + \sigma_{xy}^2)}{(2 + Z^{\textrm{H}} \sigma_{xx}) (2 + Z^{\textrm{E}} \sigma_{xx}) + \eta_0^2 \sigma_{xy}^2} \,,
\label{rTMTM}
\end{equation}

\noindent where $Z^{\textrm{E}} = k_{0z}/(\omega_0 \epsilon_0)$, $Z^{\textrm{H}} = \omega_0 \mu_0/k_{0z}$, $\eta_0^2 = \mu_0/\epsilon_0$ and $\mu_0$ is the magnetic permeability of vacuum. Here, we shall neglect spatial dispersion and the expressions to be used for the longitudinal and transverse conductivities were obtained in Ref.~\cite{Gusynin2007} from an approach in the quantum context applying the Kubo formula, yielding
\begin{widetext}
\begin{align}
	\sigma_{xx} (\omega, B) = \frac{e^3 v_F^2 B \hbar (\omega + i \tau^{-1})}{i \pi} \sum_{n = 0}^{\infty} \bigg\{ &\frac{n_F (M_n) - n_F (M_{n+1}) + n_F (- M_{n+1}) - n_F (- M_n)}{(M_{n+1} - M_n) \left[ (M_{n+1} - M_n)^2 - \hbar^2 (\omega + i \tau^{-1})^2 \right]} \nonumber \\
	+ &\frac{n_F (- M_n) - n_F (M_{n+1}) + n_F (- M_{n+1}) - n_F (M_n)}{(M_{n+1} + M_n) \left[ (M_{n+1} + M_n)^2 - \hbar^2 (\omega + i \tau^{-1})^2 \right]} \bigg\} \,,
\label{sigmaxx}
\end{align}
\vspace{-0.5cm}
\begin{align}
	\sigma_{xy} (\omega, B) = - \frac{e^3 v_F^2 B}{\pi} \sum_{n = 0}^{\infty} &\left[ n_F (M_n) - n_F (M_{n+1}) - n_F (- M_{n+1}) + n_F (- M_n) \right] \nonumber \\
	\times &\left[ \frac{1}{(M_{n+1} - M_n)^2 - \hbar^2 (\omega + i \tau^{-1})^2} + \frac{1}{(M_{n+1} + M_n)^2 - \hbar^2 (\omega + i \tau^{-1})^2} \right] \,.
\label{sigmaxy}
\end{align}
\end{widetext}

\noindent Due to the magnetic field, the graphene energy spectrum is quantized into nonequidistant Landau levels (LLs), with energies given by $M_n = \textrm{sign}(n)\sqrt{2 |n| \hbar v_F^2 e B}$, where $n = 0, \pm 1, \pm 2, ...$, $v_F = 10^6$~m/s is the Fermi velocity and $- e$ is the electron charge \cite{Gusynin2007}. Also,  $n_F (E) = [1 + e^{(E - \mu_c)/k_B T}]^{-1}$ is the Fermi-Dirac distribution, $\mu_c$ is the chemical potential and $\tau^{-1}$ is a phenomenological scattering rate which causes a small broadening in the LLs (throughout this paper we shall take $\tau = 1$~ps).

\begin{figure*}
\begin{center}
\includegraphics[width=17.8cm]{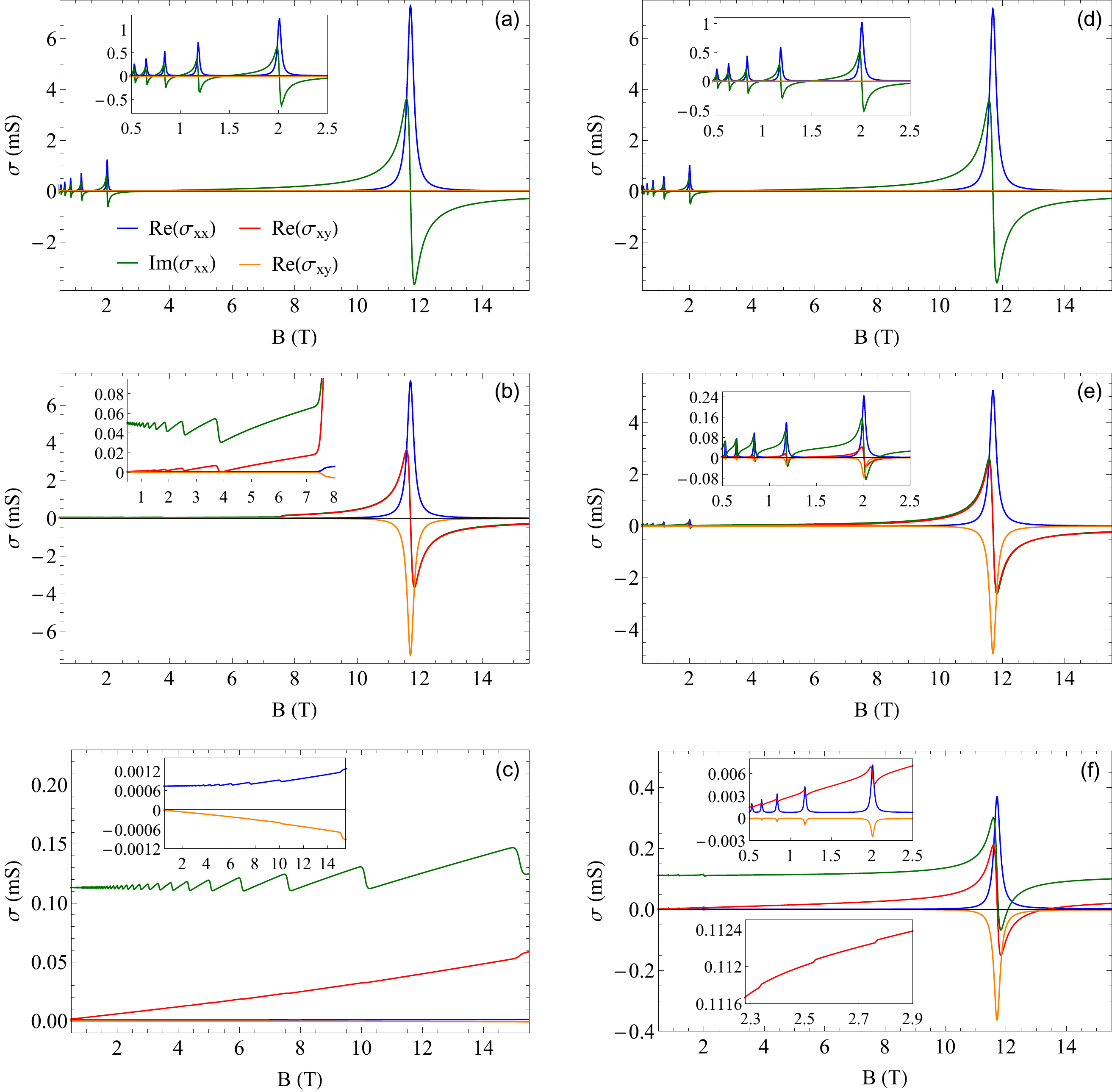}
\end{center}
\vskip -0.5cm
\caption{Real and imaginary parts of the longitudinal and transverse conductivities of graphene as functions of the external magnetic field for $\omega_0 = 6 \pi \times 10^{13}$~rad/s, $v_F = 10^6$~m/s and $\tau = 1$~ps. The first, second and third rows were obtained using $\mu_c = 0$~eV, $\mu_c = 0.1$~eV and $\mu_c = 0.2$~eV, respectively. Also, the first column was evaluated with $T = 4$~K while the second one, with $T = 300$~K.}
\label{condmuT}
\end{figure*}

From Eqs. (\ref{sigmaxx}) and (\ref{sigmaxy}), one can see that these conductivities are quite sensitive to variations in some parameters. In particular, the density of the charge carriers depends heavily on the temperature of the medium, so that, in order to explore its effect on the RET rate, we analyze the conductivities at low and room temperatures. Figure \ref{condmuT} portrays the real and imaginary parts of the longitudinal and transverse conductivities as functions of the external magnetic field $B$. Each row shows the behavior for a different value of chemical potential $\mu_c$ ($0$~eV, $0.1$~eV and $0.2$~eV, respectively). Panels (a)-(c) illustrate the behavior for temperature $T~=~4$~K, whilst (d)-(f) are results for $T~=~300$~K. In all of them, we consider $\omega_0 = 6 \pi \times 10^{13}$~rad/s ($\lambda_0 = 2 \pi c/ \omega_0 = 10$~$\mu$m) and intensities of $B < 16$~T. The dependence with $B$ is not simple, so let us begin with Fig. \ref{condmuT}(a). The sharp peaks appear whenever $\hbar \omega_0$ equals the difference in energy between two LLs whose intraband or interband transition is allowed by selection rules and the Fermi-Dirac distribution (which, in this case of low temperature, resembles a step function). For instance, the largest peak around $B \approx 11.6$~T is due to the resonance of $\hbar \omega_0$ with the first intraband transition ($0 \rightarrow 1$), while the others are due to interband transitions ($- n \rightarrow n + 1$, $- n - 1 \rightarrow n$). Despite being vanishingly small, as expected from Eq. (\ref{sigmaxy}), we plotted the transverse conductivity for $\mu_c = 0$~eV for consistency. In Figs. \ref{condmuT}(b) and \ref{condmuT}(c), we have $\mu_c \neq 0$ and a feature that stands out are the discontinuities in the plots. As $B$ increases, the LLs also increase in energy and these discontinuities show up each time a given LL crosses the chemical potential value. They occur whenever $M_n = \mu_c$, so that the corresponding value of the magnetic field is obtained from
\begin{equation}
	B = \frac{\mu_c^2}{2 n \hbar e v_F^2} \,,
\end{equation}

\noindent valid for $n > 0$. In the case of Fig. \ref{condmuT}(b) ($\mu_c = 0.1$~eV), the crossing of the last LL ($n = 1$) occurs for $B \approx 7.6$~T. This explains why we can still see the sharp peak around $B \approx 11.6$~T that is generated from the resonance of $\hbar \omega_0$ with the intraband transition $0 \rightarrow 1$, since $M_0 < \mu_c < M_1$ for such region of field intensities. On the other hand, resonances with smaller $B$ do not appear in this plot because these interband transitions are never allowed by the Fermi-Dirac distribution. The most extreme case is seen in Fig. \ref{condmuT}(c), in which no transition between LLs contributes and only discontinuities take place.

We now switch to the results at room temperature (the second column of panels in Fig. \ref{condmuT}). In short, the mathematical outcome of increasing the temperature is to provide longer decay tails to the Fermi-Dirac distribution of graphene. As an immediate consequence, more LLs are allowed to have a non zero occupation probability and, hence, new contributions from multiple transitions between LLs can emerge because of the thermal fluctuations. So where there were solely effects of the discontinuities, we now notice the two intertwined key features previously reported: {\it (i)} the sharp peaks due to the resonances of $\hbar \omega_0$ and {\it (ii)} the discontinuities arising from the crossings, but smoothed by the higher temperature and appearing as small steps as shown in the bottom inset of Fig. \ref{condmuT}(f). They can also be seen in the curves of (e) if we zoom in enough. However they do not exist in (d) as it is the case of zero chemical potential and, consequently, there are no crossings of the LLs [this result is very similar to the one obtained in (a)]. The positions of the peaks mentioned in {\it (i)} are independent of $\mu_c$ and $T$, so that they always manifest themselves at the same values of $B$ in all the curves of Fig. \ref{condmuT}. In the case of larger values of the chemical potential, combined with the smooth profile of the Fermi-Dirac distribution at $T = 300$~K, even higher peaks for a few of the subsequent intraband transitions ($1 \rightarrow 2$, $2 \rightarrow 3$) are allowed, but they happen at somewhat unrealistic values of the magnetic field around $68.2$~T and $115.8$~T, and therefore are not shown in the plots. Incidentally, this explains why the curves in Fig. \ref{condmuT}(f) do not go to zero after the peak, that is the effect of the $1 \rightarrow 2$ transition kicking in.


\section{\label{SecResults}Results and discussions}


\begin{figure*}
\begin{center}
\includegraphics[width=17.8cm]{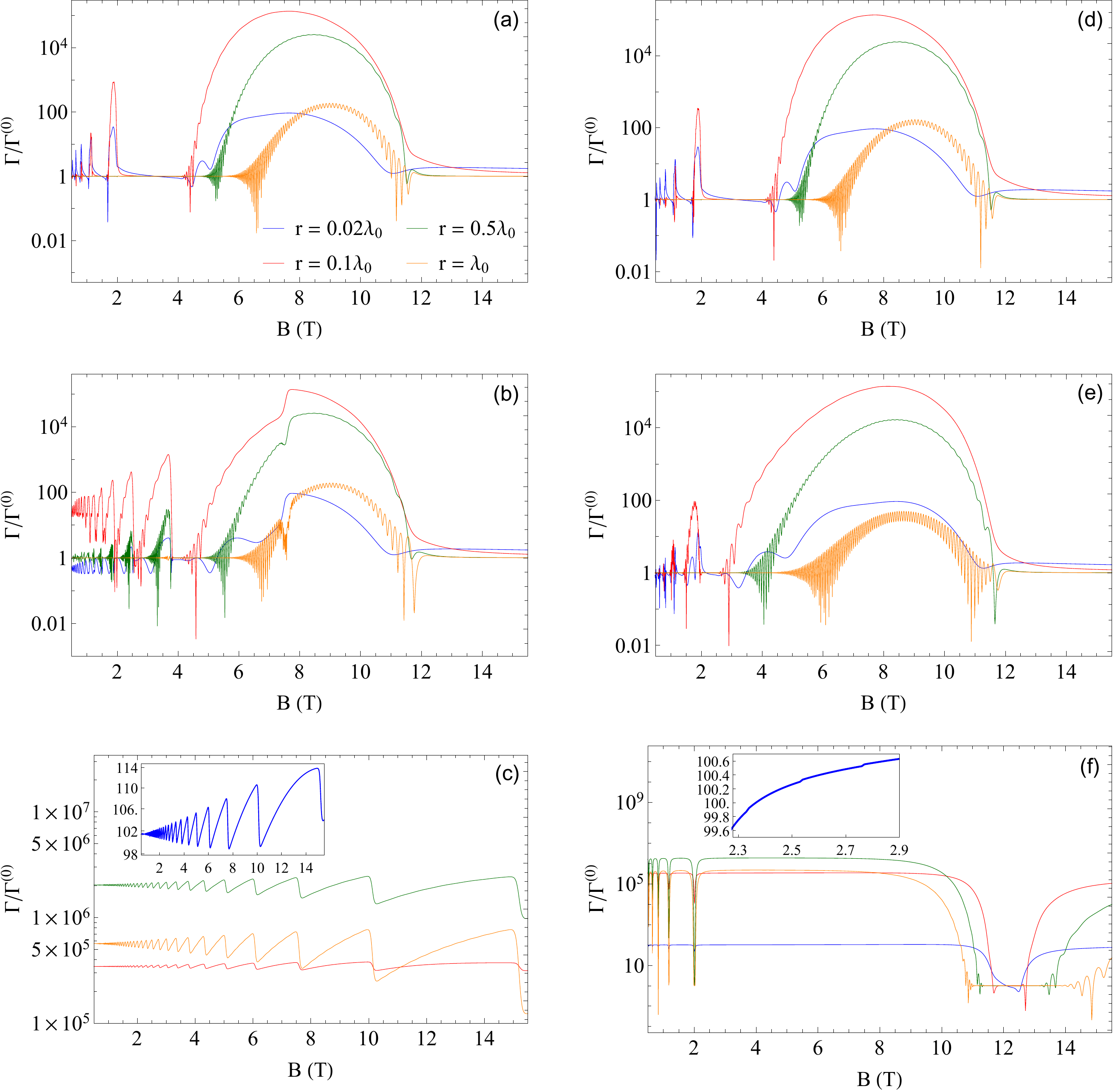}
\end{center}
\vskip -0.5cm
\caption{Normalized RET rate as functions of the external magnetic field. Each color represents a separation $r$ between the emitters with dominant transition wavelength $\lambda_0 = 10$~$\mu$m, both at a distance $z = 50$~nm from the graphene sheet. The first, second and third row panels were obtained using $\mu_c = 0$~eV, $\mu_c = 0.1$~eV and $\mu_c = 0.2$~eV, respectively. Also, the first column was evaluated with $T = 4$~K while the second shows results for $T = 300$~K.}
\label{RETmuTz0005}
\end{figure*}

The results for the resonance energy transfer were evaluated using the same parameters presented in the analysis of the conductivities. Figure \ref{RETmuTz0005} depicts the normalized RET rate calculated according to Eq. (\ref{RETRateSimpl}) as a function of the applied magnetic field for four different configurations of distance $r$ between the emitters. Panels (a)-(c) and (d)-(f) refer to temperatures $T~=~4$~K and $T~=~300$~K, respectively, and each row refers to a chemical potential value exactly as in Fig. \ref{condmuT}. We chose to work in the near-field region ($z~=~50$~nm~$\ll~\lambda_0$) in order to explore the interaction of the emitters with the graphene's surface magnetoplasmon polaritons (MPPs), as we shall elaborate later on. One could expect a Zeeman splitting for the values of $B$ considered here, as well as a $z$-dependent Casimir shift of the emitters transition energy. However, in the unlikely event that such effects do significantly shift the ``bare" frequency $\omega_0$ (the electric and/or magnetic polarizabilities of the emitters would have to be abnormally large), it would be just a matter of replacing the shifted frequency in our calculations. 

It should be noticed that the results for the normalized RET rate in Fig. \ref{RETmuTz0005} are naturally correlated with the response of graphene to the external field, expressed in terms of its longitudinal and transverse conductivities. In this sense, when the magnetic field gets close to a value for which the conductivities present a discontinuity (whose reason was discussed in Sec. \ref{SubsecGraphProp}), this effect is directly reflected in the RET rate. Analogously, whenever there is a contribution coming from permitted transitions between LLs, the normalized RET rate is drastically reduced and then increases again while there are still magnetic field values to which other permitted transitions may contribute.

From the plots of Fig. \ref{RETmuTz0005}, a key fact that stands out is a striking non-monotonic dependence on $r$. When the emitters are very close to each other, the excitation transfer is dominated by the free-space channel, and the graphene impact is not so significant. By increasing $r$ (and keeping $z$ fixed), the environment starts to play a more important role and the relative RET rate shoots up by orders of magnitude. Finally, by increasing $r$ even more, the maximum of $\Gamma/\Gamma^{(0)}$ shrinks about 2 orders of magnitude for $\mu_c = 0, 0.1$~eV and drops about a factor of 10 for $\mu_c = 0.2$~eV. Such effect occurs in a similar way for both temperatures studied.

In order to explain such an impressive variation of the RET rate, it is necessary to make a small digression about the graphene mode structure and, in particular, of its MPPs. The MPPs are surface waves allowed by Maxwell equations under certain boundary conditions. Such surface waves are characterized by the decaying behavior in the $z$-direction in both sides of the graphene sheet and they must be associated with a pole in the reflection coefficients \cite{NunoBook}. Therefore, from Eq. (\ref{rTMTM}) we have
\begin{equation}
	(2 + Z^{\textrm{H}} \sigma_{xx}) (2 + Z^{\textrm{E}} \sigma_{xx}) + \eta_0^2 \sigma_{xy}^2 = 0 \,.
\end{equation}

\noindent Enforcing a relation between $k_{\|}$ and $\omega_0$, we arrive at the general dispersion relation for the MPPs \cite{NunoBook}. A straightforward manipulation gives
\begin{widetext}
\begin{equation}
	k_{\|}^4 + \frac{4 \omega_0^2}{c^2} \left\{ \frac{1}{\eta_0^2 \sigma_{xx}^2} \left[1 + \frac{\eta_0^2}{4} \left( \sigma_{xx}^2 + \sigma_{xy}^2 \right) \right]^2 - 1 \right\} k_{\|}^2 - \frac{4 \omega_0^4}{c^4} \left\{ \frac{1}{\eta_0^2 \sigma_{xx}^2} \left[1 + \frac{\eta_0^2}{4} \left( \sigma_{xx}^2 + \sigma_{xy}^2 \right) \right]^2 - 1 \right\} = 0 \,,
\end{equation}

\noindent leading to
\begin{equation}
	k_{\|}^2 = \frac{2 \omega_0^2}{c^2} \left\{ 1 - \frac{1}{\eta_0^2 \sigma_{xx}^2}\left[1 + \frac{\eta_0^2}{4} \left( \sigma_{xx}^2 + \sigma_{xy}^2 \right) \right]^2 \right\} \left[1 \mp \sqrt{1 + \frac{\eta_0^2 \sigma_{xx}^2}{1 - \dfrac{\eta_0^2}{2} \left( \sigma_{xx}^2 - \sigma_{xy}^2 \right) + \dfrac{\eta_0^4}{16} \left( \sigma_{xx}^2 + \sigma_{xy}^2 \right)^2} } \right] \,.
\end{equation}
\end{widetext}

\noindent The solutions that interest us are those whose real part of $k_{\|}$ is positive \cite{NunoBook}. In order to handle with the previous relation, we can use the fact that, away from the intense variation around $B = 11.6$~T, we have $\eta_0^2 \sigma_{xx}^2 \ll 1$ and also $\eta_0^2 \sigma_{xy}^2 \ll 1$. Hence, it is reasonable to expand this formula and retain only its first terms, yielding 
\begin{align}
	k_{\|}^{(+)} &= k_{\textrm{MPP}} \approx \frac{2 i \epsilon_0 \omega_0}{\sigma_{xx}} \,, 
\label{kpMPP} \\
	k_{\|}^{(-)} &= k_{\textrm{QTE}} \approx \frac{\omega_0}{c} \sqrt{1 - \frac{\eta_0^4}{4} \left (\sigma_{xx}^2 - \sigma_{xy}^2 \right)^2} \,.
\label{kpQTE}
\end{align}

\noindent The so called quasi-transverse-electric (QTE) modes \cite{Ferreira2012} given by (\ref{kpQTE}) play virtually no role in the RET, while the MPP branch (\ref{kpMPP}) is the main focus of this work. The fact that $k_{\textrm{MPP}}$ is not purely real indicates that such surface modes have a dissipative character and therefore a finite propagation length {\it parallel} to the graphene's surface \cite{NunoBook}, given roughly by
\begin{equation}
	L_{\textrm{MPP}} \approx \frac{1}{{\rm Im}(k_{\textrm{MPP}})} = \frac{1}{2 \epsilon_0 \omega_0} \frac{|\sigma_{xx}|^2}{{\rm Re} \, \sigma_{xx}} \,.
\label{LMPP}
\end{equation}

\begin{figure}
\begin{center}
\includegraphics[width=7.8cm]{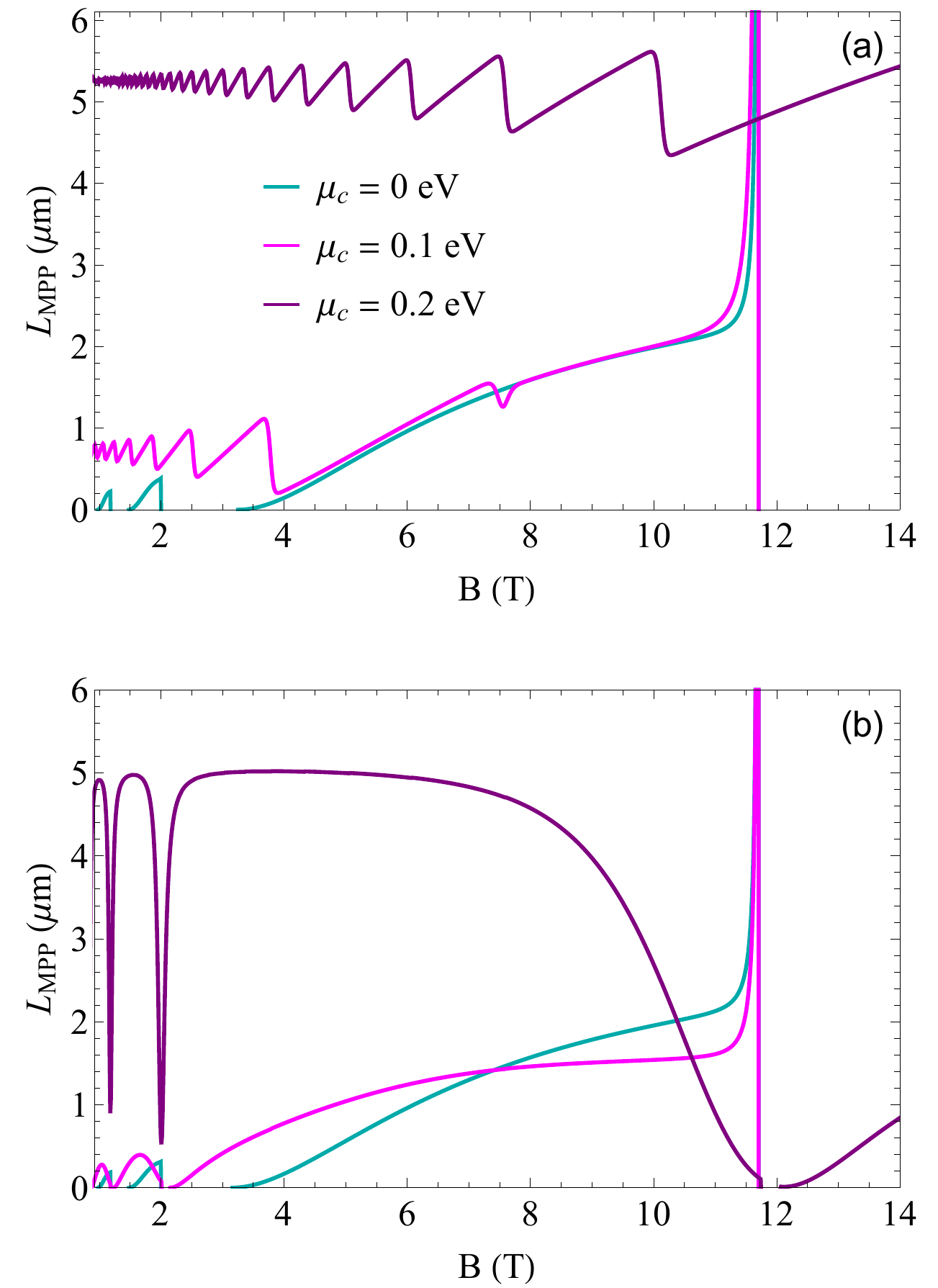}
\end{center}
\vskip -0.5cm
\caption{MPP propagation length as a function of the magnetic field for different values of the chemical potential and {\bf (a)} $T = 4$~K and {\bf (b)} $T = 300$~K. The same parameters used in the analysis of the conductivities were also employed here.}
\label{LMPPmu00102T}
\end{figure}

In Fig. \ref{LMPPmu00102T}, the propagation length of the MPPs is plotted as a function of the external magnetic field for the three values of chemical potential considered before. The upper and lower plots correspond to calculations using $T~=~4$~K and $T~=~300$~K, respectively, and, in broad strokes, their main features can be traced back to the longitudinal conductivity. For $\mu_c = 0.2$~eV, the role of the magnetoplasmons is quite evident: we see that the two emitters are within the MPPs range for $r \lesssim 5$ $\mu$m $\approx 0.5 \lambda_0$, that explains the consistent dominance of the green curve in Figs. \ref{RETmuTz0005}(c) and \ref{RETmuTz0005}(f). It also explains the characteristic discontinuities for temperature $T = 4$~K and why there are such precipitous drops at the resonances in the case of $T = 300$~K (both clearly correlated with the results of $L_{\rm MPP}$). A similar reasoning can be extended to the set of parameters $\mu_c = 0.1$~eV and $T = 4$~K, specially for low fields, where we can also note that the two emitters are within the MPPs range for $r \lesssim 1$ $\mu$m $\approx 0.1 \lambda_0$, in agreement with the enhanced normalized RET rate obtained in Fig. \ref{RETmuTz0005} in this same configuration. This explanation is less evident for the other results of Fig. \ref{LMPPmu00102T}, but it is clear that, at least in the $B = 3-11$~T range, the steady rise in the $L_{\rm MPP}$ corresponds to the ``great hill" profile in the RET plots centered in $B \approx 8$~T. In addition, let us note that the lower $L_{\rm MPP}$ values for $\mu_c = 0, 0.1$~eV also explain the fact that the maximum relative RET occurs for shorter distances in these cases (the green ``hill" is well below the red one in panels (a), (b), (d) and (e) of Fig. \ref{RETmuTz0005}). Finally, as the distance between the emitters gets too large, they evade the propagation range of the MPPs, explaining the downard trend for $r \gtrsim L_{\rm MPP}$ in all curves of Fig. \ref{RETmuTz0005}.

A remarkable feature present in Fig. \ref{RETmuTz0005} that is still to be discussed is the extreme sensitivity of the normalized RET rate with respect to variations in the magnetic field. Indeed, we see that for $T = 300$~K, $\mu_c = 0.2$~eV and $r =\lambda_0$, the relative RET rate can change by impressive five to six orders of magnitude, even for tiny variations of magnetic field around $1$~T. We see that, by using the magnetic field as a ``dial" to tune the transition frequency to a possible LL transition, one could essentially ``turn off"  the graphene sheet, at least with respect to the RET process. Such incredible sensitivity may be also traced to fact that the MPPs depend critically upon Re $\sigma_{xx}$, so small variations in the conductivity can generate big effects in the $L_{\rm MPP}$ and huge modifications in the RET rate. As an aside, we should point out that the normalized RET inherit the small steps that are present in the conductivites, as shown in the inset of Fig. \ref{RETmuTz0005}(f).

Another interesting feature of the normalized RET rate is the oscillatory character - quite intense, for some parameters - as a function of the magnetic field. Although the previous formulas hold in all distance regimes, from now on we shall be concerned with the analysis solely in the near-field region ($\omega_0 z/c \ll 1$) in order to understand this intriguing behavior. Splitting the contribution of the propagating and evanescent modes in (\ref{GSzz}), we can write
\begin{align}
	\mathds{G}^{(\textrm{S})}_{zz} &= \frac{i c^2}{4\pi \omega_0^2} \left\{ \int_0^{\omega_0/c} \!\!\! dk_{\|} \frac{k_{\|}^3 \, J_0 (k_{\|} r) \,r^{\textrm{TM,TM}} \, e^{2i k_{0z} z}}{k_{0z}} \right.\nonumber \\
	&\left. + \int_{\omega_0/c}^\infty \!\!\! dk_{\|} \frac{k_{\|}^3 \, J_0 (k_{\|} r) \,r^{\textrm{TM,TM}} \, e^{- 2 \kappa_{0z} z}}{i \kappa_{0z}} \right\} \,,
\label{GSzzPropEva}
\end{align}

\noindent with $\kappa_{0z} = i k_{0z} = \sqrt{k_{\|}^2 - \omega_0^2/c^2}$. The evanescent part largely dominates the propagating one in the near-field regime, so Eq. (\ref{GSzzPropEva}) can be approximated to
\begin{equation}
	\mathds{G}^{(\textrm{S})}_{zz} \approx \frac{c^2}{4\pi \omega_0^2} \int_0^\infty dk_{\|} k_{\|}^2 \, J_0 (k_{\|} r) \,r^{\textrm{TM,TM}} \, e^{- 2 k_{\|} z} \,,
\label{GSzzNF}
\end{equation}

\noindent where we used $\kappa_{0z} \approx k_{\|}$. Applying the same considerations to the reflection coefficient (\ref{rTMTM}), we get
\begin{equation}
	r^{\textrm{TM,TM}} \approx \frac{k_{\|} - \dfrac{i  \eta_0 \sigma_{xx}}{2} \! \left[ 1 + \dfrac{\sigma_{xy}^2}{\sigma_{xx}^2} \right] \! \dfrac{\omega_0}{c} }{k_{\|} - i \left\{ \dfrac{2 \epsilon_0 \omega_0}{\sigma_{xx}} + \dfrac{ \eta_0 \sigma_{xx}}{2} \! \left[ 1 + \dfrac{\sigma_{xy}^2}{\sigma_{xx}^2} \right] \! \dfrac{\omega_0}{c} \right\} } \,.
\label{rTMTMNF}
\end{equation}

\noindent Moreover, away from $B \approx 11.6$~T we may retain only the very first contribution in $\eta_0 \sigma_{xx}$, yielding 
\begin{align}
	r^{\textrm{TM,TM}} &\approx \frac{k_{\|}}{k_{\|} - \dfrac{2 i \epsilon_0 \omega_0}{\sigma_{xx}} } \,,
\label{rTMTMNF}
\end{align}

\noindent from which one immediately identifies the magnetoplasmon polariton at the pole $k_{\textrm{MPP}}~=~2 i \epsilon_0 \omega_0 / \sigma_{xx}$ in accordance with the result obtained in Eq. (\ref{kpMPP}). The substitution of Eq. (\ref{rTMTMNF}) in Eq. (\ref{GSzzNF}) leads us to a simpler expression for the scattering Green function, to wit
\begin{equation}
	\mathds{G}^{(\textrm{S})}_{zz} \approx \frac{c^2}{4\pi \omega_0^2} \int_0^\infty dk_{\|} \frac{k_{\|}^3 \, J_0 (k_{\|} r)}{k_{\|} - \dfrac{2 i \epsilon_0 \omega_0}{\sigma_{xx}}}  \, e^{- 2 k_{\|} z} \,.
\label{GSzzNF1}
\end{equation}

\noindent Despite its relative simplicity, we could not solve (\ref{GSzzNF1}) in terms of well known functions. We are, however, particularly interested in the $|2 i  \epsilon_0 \omega_0 /\sigma_{xx}| \gg 1/z$ regime, corresponding to low magnetic fields (away from the abrupt changes at the LL transitions). Then, an analytical solution for Eq. (\ref{GSzzNF1}) is available, and also taking into account that $\textrm{Im}(\sigma_{xx}) \gg \textrm{Re}(\sigma_{xx})$, we get
\begin{align}
	\mathds{G}^{(\textrm{S})}_{zz} &\approx \frac{c^2 \, \textrm{Im}(\sigma_{xx})}{4 \pi \epsilon_0 \omega_0^3} \frac{3 z (3 r^2 - 8z^2)}{(r^2 + 4z^2)^{7/2}} \,.
\label{GSzzNF2}
\end{align}

\begin{figure}
\begin{center}
\includegraphics[width=8.6cm]{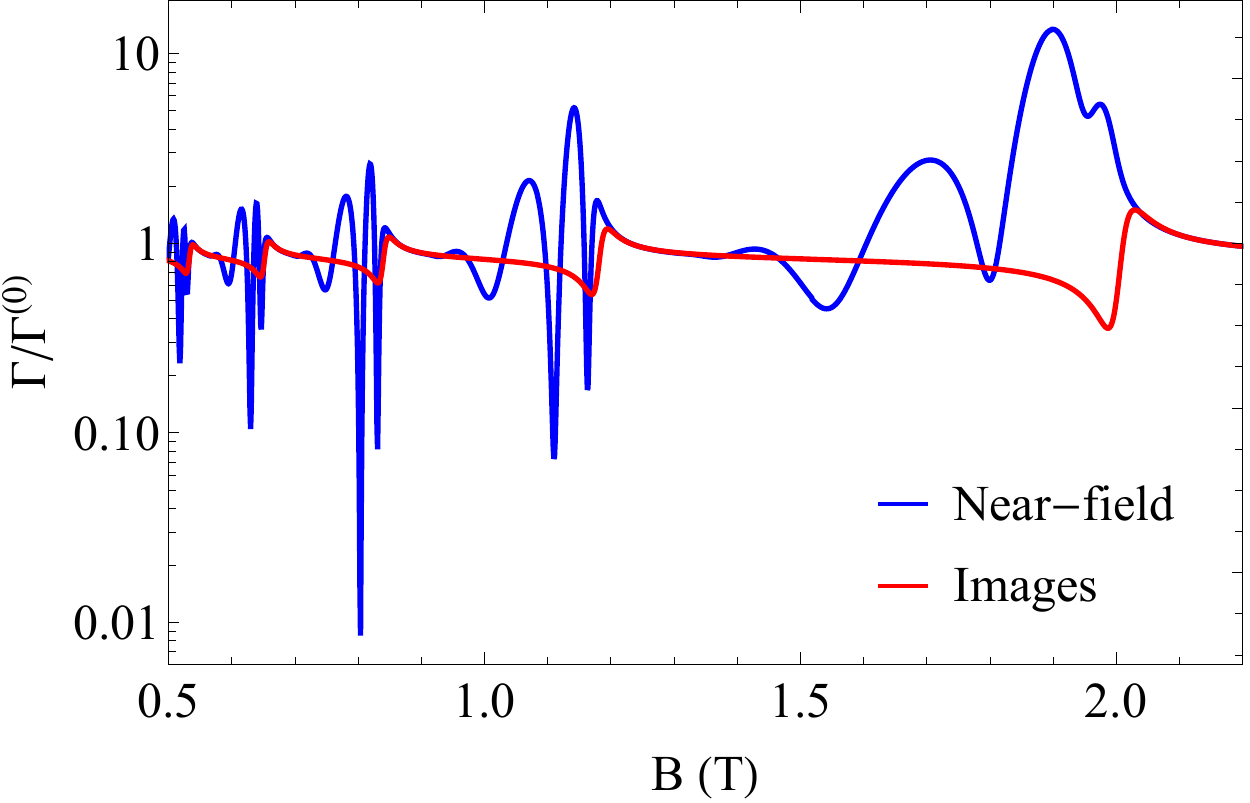}
\end{center}
\vskip -0.5cm
\caption{Normalized RET rate as a function of the magnetic field for the case previously shown with $T = 300$~K, $\mu_c = 0.1$~eV and $r~=~0.02\lambda_0$. The plots are comparisons between results obtained with $\mathds{G}^{(\textrm{S})}_{zz}$ calculated using Eqs. (\ref{GSzzNF1}) (blue curve) and (\ref{GSzzNF2}) (red curve).}
\label{RET-NF-Ima}
\end{figure}

In Fig. (\ref{RET-NF-Ima}) we depict the comparison of the RET rate using (\ref{GSzzNF1}) and (\ref{GSzzNF2}). It is clearly seen that the low field approximation captures a sort of average behavior, but fails to show the marked oscillations present in (\ref{GSzzNF1}). At this point, we remember that the denominator in Eq. (\ref{GSzzNF1}) comes from the $r^{\textrm{TM,TM}}$, whose pole provides us with the dispersion relation of the MPPs. To derive Eq. (\ref{GSzzNF2}) we effectively disregarded this pole and, consequently, the information on the contribution of the interaction with the surface plasmons. That led us to a result with a clear interpretation in terms of images - as $\sqrt{r^2+(2z)^2}$ is the distance between an emitter and the image of the other - but it should be recalled that such interpretation was not to be obviously expected: we are in the low conductivity regime, so these dressed images probably owe their appearance more to the plane symmetry than to the (short) distance regime.

\section{\label{SecConclusions}Final remarks and conclusions}


In summary, we have investigated the resonance energy transfer between two emitters near a graphene sheet in the presence of a constant, uniform and perpendicular magnetic field. The fundamental motivation was to take advantage of the remarkable magneto-optical properties of graphene in order to tailor and control the RET rate between the emitters. From our findings, we conclude that, in addition to providing us with a promising platform to manipulate atomic interaction through an external agent, the RET is particularly suitable to active manipulation due to its extreme sensitivity to variations of the magnetic field. We have demonstrated that the strongly confined magnetoplasmon polaritons supported by the graphene monolayer play a key role in the excitation transfer between the emitters. We stress that the RET rate can be enormously altered, suffering abrupt variations up to six orders of magnitude with respect to the free space value. Moreover, specially in the case of room temperature, these huge variations occur for feasible values of the magnetic field (of the order of $1$~T for appropriate choices of the system parameters), being within the scope of experimental realization. As a matter of fact, the RET modulation is so large and so sharp that magnetoactive materials could be thought as an energy transfer switch, that can be turned on and off with no physical contact. Altogether we expect that these results will not only allow for an alternative way to control the resonance energy transfer but also pave the way for the development of new devices in plasmonics and nanophotonics.


\begin{acknowledgments}

P. P. A and C. F. thank L. Martín-Moreno for enlightening discussions. C.F. and F.S.S.R. acknowledge Conselho Nacional de Desenvolvimento Científico e Tecnológico (CNPq) for financial support (grant numbers 310365/2018-0 9 and 309622/2018-2). F.S.S.R. (grant number E26/203.300/2017) and P.P.A. acknowledge Fundação de Amparo à Pesquisa do Estado do Rio de Janeiro (FAPERJ).

\end{acknowledgments}



\end{document}